\documentclass{emulateapj}

\shorttitle{Radial Angular Momentum Transfer and Magnetic Barrier for SGRB Central Engine Activity}
\shortauthors{Liu et al.} \slugcomment{}
\usepackage{textcomp}
\usepackage{amsmath}
\usepackage{graphicx}
\begin{document}
\title{Radial Angular Momentum Transfer and Magnetic Barrier for Short-Type Gamma-Ray Burst Central Engine Activity}
\author{Tong Liu\altaffilmark{1}, En-Wei Liang\altaffilmark{2,3}, Wei-Min Gu\altaffilmark{1,4}, Shu-Jin Hou\altaffilmark{1}, Wei-Hua Lei\altaffilmark{5,6}, Lin Lin\altaffilmark{3}, Zi-Gao Dai\altaffilmark{7}, and Shuang-Nan Zhang\altaffilmark{3,8,9}}

\altaffiltext{1}{Department of Physics and Institute of Theoretical Physics and Astrophysics, Xiamen University, Xiamen, Fujian 361005, China}

\altaffiltext{2}{Department of Physics and GXU-NAOC Center for Astrophysics and Space Sciences, Guangxi University, Nanning, Guangxi 530004, China; lew@gxu.edu.cn}

\altaffiltext{3}{National Astronomical Observatories, Chinese Academy of Sciences, Beijing 100012, China}

\altaffiltext{4}{Harvard-Smithsonian Center for Astrophysics, 60 Garden Street, Cambridge, MA 02138, USA}

\altaffiltext{5}{School of Physics, Huazhong University of Science and Technology, Wuhan, Hubei 430074, China}

\altaffiltext{6}{Department of Physics and Astronomy, University of Nevada Las Vegas, Las Vegas, NV 89154, USA}

\altaffiltext{7}{Department of Astronomy, Nanjing University, Nanjing, Jiangsu 210093, China}

\altaffiltext{8}{Key Laboratory of Particle Astrophysics, Institute of High Energy Physics, Chinese Academy of Sciences, Beijing 100049, China}

\altaffiltext{9}{Physics Department, University of Alabama in Huntsville, Huntsville, AL 35899, USA}

\begin{abstract}
Soft extended emission (EE) following initial hard spikes up to 100 seconds was observed with {\em Swift}/BAT for about half of short-type gamma-ray bursts (SGRBs). This challenges the conversional central engine models of SGRBs, i.e., compact star merger models. In the framework of the black hole-neutron star merger models, we study the roles of the radial angular momentum transfer in the disk and the magnetic barrier around the black hole for the activity of SGRB central engines. We show that the radial angular momentum transfer may significantly prolong the lifetime of the accretion process and multiple episodes may be switched by the magnetic barrier. Our numerical calculations based on the models of the neutrino-dominated accretion flows suggest that the disk mass is critical for producing the observed EE. In case of the mass being $\sim 0.8M_{\odot}$, our model can reproduce the observed timescale and luminosity of both the main and EE episodes in a reasonable parameter set. The predicted luminosity of the EE component is lower than the observed EE with about one order of magnitude and the timescale is shorter than 20 seconds if the disk mass being $\sim 0.2M_{\odot}$. {\em Swift}/BAT-like instruments may be not sensitive enough to detect the EE component in this case.  We argue that the EE component would be a probe for merger process and disk formation for compact star mergers.
\end{abstract}


\section{Introduction}
Gamma-ray bursts (GRBs) are sorted into two classes (Kouveliotou et al. 1993), i.e., short duration ($T_{90} < 2 \rm s$, SGRBs) and long duration GRBs ($T_{90} > 2 \rm s$). Their progenitors are thought to be mergers of two compact stars (Eichler et al. 1989; Paczy\'{n}ski 1991; Narayan et al. 1992; Zhang et al. 2007; Nakar 2007) and collapses of massive stars (Woosley 1993; Paczy\'nski 1998; Piran 2004; Zhang \& M\'esz\'aros 2004; Woosley \& Bloom 2006), respectively. However, the observed burst duration is instrumentally dependent (Donaghy et al. 2006; Qin et al. 2012). {\em Swift} observations reveal that the short-long GRB classification scheme does not always match their physical origin classification scheme, $i.e.$, mergers of compact binaries (Type I) $vs.$ collapsars (Type II) (Zhang 2006; Zhang et al. 2007; Zhang et al. 2009; L\"{u} et al. 2010; Xin et al. 2011). With the {\em CGRO}/BATSE data, Lazzati et al. (2001) found an excess emission peaking $\sim 30$ s after the prompt one, which is detectable for $\sim 100$ s for some SGRBs (see also Connaughton 2002; Norris et al. 2010). About half of the lightcurves of those GRBs that are recognized as Type I GRBs with {\em Swift}/BAT show initial hard spikes following by an extended emission (EE) component of soft gamma-rays up to $\sim 100$ $\rm s$ post the BAT trigger (Barthelmy et al. 2005; Lin et al. 2008; Perley et al. 2009; Zhang et al. 2009). The most prominent case is GRB 060614. Its lightcurve is composed of some initial hard spikes and a long, soft gamma-ray tail, which leads to its $T_{90}$ being $\sim 110$ $\rm s$ (Gehrels et al. 2006). The initial hard spikes could be recognized as a SGRB with {\em CGRO}/BATSE-like instruments, since the soft EE is out of the instrument bands (Zhang et al. 2007). No accompanied supernovae was detected for this nearby long GRB (the redshift $z=0.1254$; Della Valle et al. 2006; Fynbo et al. 2006; Gal-Yam et al. 2006), disfavoring the collapse of a massive star as the progenitor of this GRB. On the other hand, it is shown that some intrinsically short GRBs are likely of Type II origin (Zhang et al. 2009; Belczynski et al. 2010; Levesque et al. 2010; Lin et al. 2010; Xin et al. 2011; Virgili et al. 2011). These observations indicate that $T_{90}$ may be not a good parameter to distinguish two types of GRBs. A detailed analysis on the instrumental selection effect and energy dependence of $T_{90}$ with {\em Fermi}/GBM data by Qin et al. (2012) further supports this idea. L\"{u} et al. (2010) proposed a new classification parameter, i.e., $\varepsilon \equiv E_{\rm iso}/E_{\rm p,z}^{1.7}$, to group an observed GRB into the physically-motivated Type I/II classification scheme\footnote{Similarly, Goldstein et al.(2010) used the ratio of gamma-ray fluence to $E_{\rm p}$ to make GRB classification.}, where $E_{\rm iso}$ is the isotropic gamma-ray energy and $E_{\rm p, z}$ is the peak energy of the $\nu f_{\nu}$ spectrum in the rest frame. They showed that some SGRBs are sorted into the high-$\varepsilon$ group as typical Type II GRBs.

The observed EE component challenges not only the short-long GRB classification scheme, but also the conventional central engine models for SGRBs. Popular central engine models of Type I GRBs are related to the accretion on to a central compact objects that is formed from merger of a stellar compact binary, namely neutrino-dominated accretion flows (NDAFs, e.g., Popham et al. 1999; Narayan et al. 2001; Di Matteo et al. 2002; Kohri \& Mineshige 2002; Kohri et al. 2005; Gu et al. 2006; Chen \& Beloborodov 2007; Liu et al. 2007, 2008, 2010a, 2010b, 2012; Lei et al. 2009; Sun et al. 2012). Merger of a black hole (BH)-neutron star (NS) binary is the most favorite scenario. Such a system would result in a rotating BH with several solar masses surrounding by a neutrino-cooled disk. The detection of the EE component likely suggests that the central engine is not died out rapidly. Several lines of evidence from {\em Swift}/BAT observations also support this idea. It was also proposed that the early shallow decay X-ray emission and internal X-ray plateau may be due to the spin-down energy release of the proto-magnetar of a stellar compact binary merger (Dai \& Lu 1998; Zhang \& M\'esz\'aros 2002; Lyons et al. 2010). Late X-ray flares may signal the restart of the GRB central engine and evolution of the disk (Fan \& Wei 2005; King et al. 2005; Burrows et al. 2006; Dai et al. 2006; Perna et al. 2006; Proga \& Zhang 2006; Lazzati et al. 2008; Lee et al. 2009; Yuan \& Zhang 2012). Lazzati et al. (2008) investigated the temporal evolution of the disk in GRB central engines to explain the observed decline of X-ray flare luminosity. They argued that it is the dynamics of the disk or the jet launching mechanism to generate an intrinsically unsteady outflow on time-scales much longer than the dynamical timescale of the system for the late X-ray flares. It was also suggested that propagation instabilities, rather than variability in the engine luminosity, are responsible for some X-ray flares (Lazzati et al. 2011).

Different from the early shallow decay X-ray emission and internal X-ray plateau, the EE component is usually highly variable and is usually not clearly separated from the burst itself in their lightcurves. It may be produced by the same process as the prompt emission in different episodes. Metzger et al. (2008) presented time-dependent models of the remnant accretion disks created during compact object mergers. They calculated the dynamics near the outer edge of the disk to study the evolution of the accretion rate at a long timescale (100 s or longer). They showed that the late-time accretion can in principle provide sufficient energy to power the late time activity observed by {\em Swift}/BAT from some SGRBs. In their models, the majority of the disk mass is in the outer edge and the disk becomes advective at the late time. In this paper, we focus on the radial angular momentum transfer in the disk and the magnetic barrier around the BH that may affect the activity of SGRB central engines. Based on the NDAF models, we present detailed calculations and apply our model to some typical SGRBs with detection of the EE component. We describe our model in Section 2. Numerical results are shown in Section 3. Conclusions and discussion are presented in Section 4.

\section{Model}
Assuming that the progenitor of SGRBs is a BH-NS binary, the merger of this system would result in a rotating BH surrounding by a neutrino-cooled disk. We focus on the roles of the radial angular momentum transfer in the disk and the magnetic barrier for prolonging the life time of the SGRB central engine (e.g., Proga \& Zhang 2006). A fraction of the disk matter may carry part of the angular momentum of the accretion matter to form a radial outflow. The competition of the radial angular momentum transfer to the gravity of the central BH would increase the timescale of the accretion process (see, e.g., Lee \& Ramirez-Ruiz 2007; Rosswog 2007; Metzger et al. 2010). On the other hand, the magnetic field accumulated near the horizon of the BH may be strong enough to prevent the gravity and accretion process. The magnetic barrier would dissipate quickly as the accretion rate drops. These processes likely lead to multiple well-connected accretion episodes (e.g. Narayan et al. 2003; Cao 2011). We illustrate the processes of our model and corresponding cartoon lightcurve in Figure 1. Our model is elaborated below.
\subsection{Outward angular momentum transfer}
Without considering the mass (energy) and angular momentum lost in jet production, the conservations of the mass (energy) and angular momentum read (e.g., Bardeen 1970; Thorne 1974; Wang et al. 2002),
\begin{equation}
M_{\rm n+1}-M_{\rm n} = (M^\ast_{\rm n}-M^\ast_{\rm n+1}){e_{\rm in, n}} = \dot{M}_{\rm n} T_{\rm n} {e_{\rm in, n}},
\end{equation}
\begin{equation}
J_{\rm n+1}-J_{\rm n}= J^\ast_{\rm n}-J^\ast_{\rm n+1}= \dot{M}_{\rm n} T_{\rm n} {l_{\rm in, n}},
\end{equation}
where $M_{\rm n}$ ($J_{\rm n}$) and $M^\ast_{\rm n}$ ($J^\ast_{\rm n}$) are masses (angular momentums) of the BH and disk, respectively, $\dot{M}_{\rm n}$ is the mass accretion rate, $T_{\rm n}$ is the accretion timescale, and ${e_{\rm in, n}}$ and ${l_{\rm in, n}}$ are the specific energy and angular momentum at inner boundary orbit in the $n$th episode ($\rm n=1, 2, 3, ...$). The mass of the disk can be written as
\begin{equation}
M^\ast_{\rm n} = 2 \pi \int_{r_{\rm in, n}}^{r_{\rm out, n}}\Sigma_{\rm n} r dr,
\end{equation}
where $r_{\rm in, n}$ and $r_{\rm out, n}$ are the inner and outer boundaries of the disk in the $n$th episode, respectively. The angular momentums of the BH and disk can be calculated with
\begin{eqnarray}
J_{\rm n} = \frac{{a_*}_{\rm n} G M_{\rm n}^2}{c},
\end{eqnarray}
\begin{eqnarray}
J^\ast_{\rm n} = 2 \pi \int_{r_{\rm in, n}}^{r_{\rm out, n}} \Sigma_{\rm n} l_{\rm n} r dr,
\end{eqnarray}
where $l_{\rm n}$ is the specific angular momentum per unit mass. It is known that ${a_*}_{\rm n}$ cannot exceed unity (e.g., Janiuk et al. 2008). Since $M^\ast_1$ supplies all the accretion process, we have
\begin{equation}
M^\ast_1 = \sum_{\rm n} \dot{M}_{\rm n} T_{\rm n},
\end{equation}
where $T_{\rm n}$ is the timescale of the $n$th epoch, which depends on the competition between the magnetic flux pressure and the gravity to the accreting mass.

\subsection{Switch of the magnetic field}
Since the predecessor of the accretion disk is a highly-magnetic NS, the conservation of magnetic flux requires an inherited magnetic field in the disk. The successive magnetic flux from the remanent NS can be given by
\begin{equation}
\Phi_{\rm NS} = 2 \pi \int_{r_{\rm in, 1}}^{r_{\rm out, 1}} B_1 r dr,
\end{equation}
where $B_1$ is the magnetic induction strength of the disk in the initial condition. The instabilities become effective when the magnetic pressure in the radial direction can support against gravity of the BH (e.g., Spruit \& Uzdensky 2005; Proga \& Zhang 2006). The accretion process may be closed, if the magnetic induction strength satisfies a critical value $B_{\rm crit, n}$,
\begin{equation}
\frac{B_{\rm crit, n}^2}{4 \pi} \sim \frac{G M_{\rm n} \Sigma_{\rm n}|_{r_{\rm n}=r_{\rm in, n}}}{r_{\rm in, n}^2},
\end{equation}
where $M_{\rm n}$ and $\Sigma_{\rm n}$ are the mass of the BH and surface density in the $n$th episode. The critical radius $r_{\rm crit, n} (\leq r_{\rm out})$ can be estimated with
\begin{equation}
\ 2 \int_{r_{\rm in, n}}^{r_{\rm crit, n}} B_{\rm n} r dr = B_{\rm crit, n} r_{\rm in, n}^2,
\end{equation}
where $B_{\rm n}$ is calculated by the conservation of magnetic flux. Therefore, the timescale is obtained by
\begin{equation}
T_{\rm n} \sim \frac{r_{\rm n}} {\left\vert \bar{v}_{\rm n} \right\vert},
\end{equation}
where $\left\vert \bar{v}_{\rm n} \right\vert$ is the absolute value of the average radial velocity from $r_{\rm crit, n}$ to $r_{\rm in, n}$.

We estimate the magnetospheric radius from
\begin{equation}r_{\rm m} \approx 6\times10^3 (v/v_{ff})^{2/3} (\dot{M}/M_\odot ~{\rm s^{-1}})^{-2/3} (M/3M_\odot)^{-4/3} r_g,
\end{equation}
where $v$, $v_{ff}$, $\dot{M}$ and $r_g$($=2GM/c^2$) are the radial velocity of the accretion disk, the free fall velocity, the mass accretion rate and Schwarzchild radius, respectively (e.g., Narayan et al. 2003; Proga \& Zhang 2006). Thus the timescale of magnetic field dissipation can be estimated as $\sim r_{\rm m}/v_{ff}$. We simplify the magnetic field as a uniform field with a magnetic flux $10^{29}$ $\rm G~cm^2$ in our calculation (e.g., Proga \& Zhang 2006). For $M_1=3$ $M_\odot$, ${a_*}_1=0.9$, $\dot{M}_1 =0.05$ $M_\odot~\rm s^{-1}$, $\alpha_1= 0.01$ and $v/v_{ff}\sim10^{-2}-10^{-3}$, the timescale of the first episode is $T_1\sim 2$ $\rm s$ from Eq. (10), and the timescale of magnetic field dissipation is $\sim 0.1-1$ $\rm s$. If the remanent magnetic flux is reduced to $5 \times 10^{28}$ $\rm G~cm^2$ and $M_2 \sim M_1$, $\alpha_2 \sim \alpha_1$, ${a_*}_2 \sim {a_*}_1$, the timescale of the second episode is $T_2\sim 27$ s in case of $\dot{M}_2= 0.01$ $M_\odot~\rm s^{-1}$. Therefore, our model can potentially explain the EE of SGRBs.

\subsection{NDAF model}

We adopt the method present by Riffert \& Herold (1995). This method is dedicated to numerically investigate the NDAF in vicinity of a rotating BH. The method defines general relativistic correction factors quoted as below,
\begin{eqnarray}
A &=& 1-\frac{2GM}{c^2 r}+\Big(\frac{a_* G M}{c^2 r}\Big)^2, \\
B &=& 1- \frac{3GM}{c^2 r}+2a_* \Big(\frac{G M}{c^2 r}\Big)^{\frac{3}{2}},\\
C &=& 1- 4a_* \Big(\frac{G M}{c^2 r}\Big)^{\frac{3}{2}}+3\Big(\frac{a_* G M}{c^2 r}\Big)^2, \\
D &=& \int_{r_{\rm ms}}^{r} \frac{\frac{x^2 c^4}{2G^2}-\frac{3xMc^2}{G}+ 4(\frac{x a_*^2 M^3 c^2}{G})^{\frac{1}{2}}-\frac{3a_*^2 M^2}{2}}{(x r)^{\frac{1}{2}}[\frac{x^2 c^4}{G^2}-\frac{3xMc^2}{G}+2 (\frac{x a_*^2 M^3 c^2}{G})^{\frac{1}{2}}]} dx,
\end{eqnarray}
where $M$, $a_*$ and $r_{\rm ms}$ are the mass, dimensionless spin parameter and the radius of marginally stable orbit of the BH (e.g., Kato et al. 2008), respectively. The continuity equation remains valid,
\begin{equation}
\dot{M}=-2 \pi r \Sigma v.
\end{equation}
The hydrostatic equilibrium in the vertical direction leads to a corrected expression for the half thickness of the disk (Riffert \& Herold 1995; Lei et al. 2009; Liu et al. 2010b),
\begin{equation}
\ H \simeq c_{\rm s}\Big(\frac{ r^3}{G M}\Big)^{\frac{1}{2}} \Big(\frac{B}{C}\Big)^{\frac{1}{2}},
\end{equation}
where $c_{\rm s}=(p/\rho)^{1/2}$ is the isothermal sound speed, $p$ and $\rho$ are the total pressure and density of the disk, respectively. The viscous shear $T_{r \phi}$ is also corrected as (Liu et al. 2010b)
\begin{equation}
\ T_{r \phi} = - \alpha p \frac{A}{(BC)^{\frac{1}{2}}},
\end{equation}
where $\alpha$ is a dimensionless constant that absorbs all the detailed microphysics of the viscous processes. The angular momentum equation can be simplified as
\begin{equation}\ T_{r \phi} =
\frac{\dot{M}}{4\pi H }\Big(\frac{GM}{r^3}\Big)^{\frac{1}{2}} \Big(\frac{D}{A}\Big)^{\frac{1}{2}}.
\end{equation}

The total pressure includes the gas pressure from nucleons $p_{\rm gas}$, radiation pressure of photons $p_{\rm rad}$, degeneracy pressure of electrons $p_{\rm e}$, and radiation pressure of neutrinos $p_\nu$ (see e.g., Liu et al. 2007),
\begin{equation}
p = p_{\rm gas} + p_{\rm rad} + p_{\rm e} + p_\nu.
\end{equation}
The energy equation is given by
\begin{equation}
Q_{\rm vis} = Q_{\rm adv} + Q_{\rm photo} + Q_\nu,
\end{equation}
where $Q_{\rm vis}$, $Q_{\rm adv}$, $Q_{\rm photo}$ and $Q_\nu$ are the viscous heating rate, the advective cooling rate, the cooling rate due to photodisintegration of $\alpha$-particles and the cooling due to the neutrino radiation, respectively (see e.g., Liu et al. 2007). The heating rate $Q_{\rm vis}$ is expressed as
\begin{equation}
Q_{\rm vis} = \frac{3GM \dot{M}}{8 \pi r^3} \frac{D}{B}.
\end{equation}
The radiation luminosity of the neutrinos released from the disk is obtained with the neutrino cooling rate $Q_\nu$, i.e.,
\begin{equation}
\ L_{\rm \nu}=4 \pi \int_{r_{\rm in}}^{r_{\rm out}} Q_{\rm \nu} r d r,
\end{equation}

We follow the approach by Ruffert et al. (1997), Popham et al. (1999), and Rosswog et al. (2003) to calculate the neutrino annihilation luminosity. The disk is modeled as a grid of cells in the equatorial plane. A cell $k$ has its mean neutrino energy $\varepsilon_{\nu_i}^k$, neutrino radiation luminosity $l_{\nu_i}^k$, and distance to a space point above (or below) the disk $d_k$. The angle at which neutrinos from cell $k$ encounter antineutrinos from another cell $k'$ at that point is denoted as $\theta_{kk'}$. Then the neutrino annihilation luminosity at that point is given by the summation over all pairs of cells,
\begin{eqnarray}
l_{\nu_i \overline{\nu}_i}&=& A_{1} \sum_k \frac{l_{\nu_i}^k}{d_k^2} \sum_{k'} \frac{l_{\overline{\nu}_i}^{k'}}{d_{k'}^2} (\varepsilon_{\nu_i}^k + \varepsilon_{\overline{\nu}_i}^{k'}) {(1-\cos {\theta_{kk'}})}^2 \nonumber\\&+& A_{2} \sum_k \frac{l_{\nu_i}^k}{d_k^2} \sum_{k'} \frac{l_{\overline {\nu}_i}^{k'}}{d_{k'}^2} \frac{\varepsilon_{\nu_i}^k +\varepsilon_{\overline{\nu}_i}^{k'}}{\varepsilon_{\nu_i}^k \varepsilon_{\overline{\nu}_i}^{k'}} {(1-\cos {\theta_{kk'}})},
\end{eqnarray}
where $A_{1} \approx 1.7\times 10^{-44} \ \rm cm~erg^{-2}~s^{-1}$ and $A_{2} \approx 1.6\times 10^{-56} \ \rm cm~erg^{-2}~s^{-1}$ (e.g., Popham et al. 1999). The total neutrino annihilation luminosity is integrated over the whole space outside the BH and disk,
\begin{equation}\ L_{\nu \overline{\nu}}=4
\pi \sum _i \int_{r_{\rm in}}^\infty \int_H^\infty l_{\nu_i \overline{\nu}_i} r drdz.
\end{equation}
However, as shown by Popham et al. (1999) and Liu et al. (2007), the neutrino annihilation would inject a highly beaming outflow around the inner part of the disk. For $M=3$ $M_\odot$, $\dot{M}=0.01 \sim 1$ $M_\odot~\rm s^{-1}$ and $\alpha=0.01 \sim 0.1$, the opening angle of the ejection $\theta$ is about $10^\circ \sim 20^\circ$. We conservatively assume that the opening angle of ejection is $10^\circ$ and the efficiency of fireball is $\eta= 0.1$ in our calculations. The observed isotropic luminosity of the $n$th step $L_{\rm iso, n}$ can be estimated with
\begin{eqnarray}
L_{\rm iso, n}=\eta L_{\nu\bar{\nu},\rm n} /(1-\rm \cos \theta).
\end{eqnarray}

\section{Numerical Results}
Observationally, half of the Type I GRBs have EE component detection.  We show the lightcurves of some SGRBs with EE component in Figure 2. The lightcurves are visually recognized as different emission episodes. We apply our model to these SGRBs. For simplicity, we consider only two emission episodes, i.e., the initial hard spikes and the EE component. We describe our numerical method for two emission episodes as follows.

There are seven unknown variables in our model, i.e., ${a_*}_1$, ${a_*}_2$, $\dot{M}_1$, $\dot{M}_2$, $M_2$, $M^\ast_1$, and $M^\ast_2$. Therefore, a group of seven equations are required in our calculations. Three of them are from the conservation of mass (energy), i.e.,
\begin{equation}
M_2 = M_1+ \dot{M}_1 T_1 e_{\rm in, n},
\end{equation}
\begin{equation}
M^\ast_2 = M^\ast_1 - \dot{M}_1 T_1,
\end{equation}
and
\begin{equation}
M^\ast_2 = \dot{M}_2 T_2 + \delta M^\ast,
\end{equation}
where $\delta M^\ast$ is the mass of the residual disk post the second episode. Two other equations are from the conservation of angular momentum, i.e.,
\begin{equation}
J_2 = J_1 + \dot{M}_{\rm 1} T_{\rm 1} {l_{\rm in, 1}}
\end{equation}
and
\begin{equation}
J^\ast_2 = J^\ast_1 - \dot{M}_{\rm 1} T_{\rm 1} {l_{\rm in, 1}},
\end{equation}
The values of $J_1$ and $J_2$ are calculated with Eq.~(4). $J^\ast_1$ ($J^\ast_2$) is determined by $M_1 (M_2)$, ${a_*}_1 ({a_*}_2)$, $\dot{M}_1 (\dot{M}_2)$, $M^\ast_1 (M^\ast_2)$, and $\alpha$, which is calculated by Eq.~(5).  The rest two equations are related to $L_{\nu \overline{\nu}}$, which is a function of $M$, $a_*$, $\dot{M}$, and $\alpha$ in the NDAF model,
\begin{equation}
L_{{\nu\bar{\nu}},1} = f(M_1, {a_*}_1, \dot{M}_1, \alpha),
\end{equation}
\begin{equation}
L_{{\nu\bar{\nu}},2} = f(M_2, {a_*}_2, \dot{M}_2, \alpha),
\end{equation}
where the function $f$ can be obtained from Eqs.~(12)-(25). The average neutrino annihilation luminosity  in each episode is calculated assuming that the accretion rate and spin parameter of the BH are a constant. The accretion timescale in each step depends on the initial mass of the disk and the switch of the magnetic field. The procedure of our calculations is descried as following.

First, we assign the initial parameters of the BH and disk. We fix the initial mass and the spin parameter of the BH as $3 M_{\odot}$ and 0.9, respectively. The viscous parameter is assumed to be $\alpha =0.01$. The initial mass of the disk is adjustable in our calculation.

Second, we take the observed average luminosity and timescale of the initial hard spikes as $L_{\nu\bar{\nu},1}$ and $T_1$, then calculate $\dot{M}_1$ with Eq.~(32). The values of $M_2$, $M^\ast_2$, $a_{*2}$, and $\dot M_2$ are derived from Eqs.~(27), (28), (30), and (33), respectively.

Third, we calculate $T_2$ and $L_{\nu\bar{\nu},2}$ with Eqs. (6) and (33) by taking a small value of $\delta M^\ast$, i.e., about $\sim 0.1 M_\odot$.

We adjust the initial mass of the disk and find that in case of $M^\ast_1=0.8 M_\odot$ our results are roughly consistent with the observations for some typical SGRBs with EE, i.e., 050724, 060614, 061006, 061210, 070714B, 071227. Our results are shown in Figure 3. The first episodes  of these GRBs are in the left-top circle of the figure. We correspondingly derive a region of the second episodes for the circle with the universal parameters mentioned above. We notice that besides GRB 050724 and 071227, the second episodes of the other four GRBs are in the region. Note that the EE component of the two GRBs are likely a late flare, which may have different physical origin as mentioned in \S 1.

The mass of the disk is a very important factor for the EE. The decrease of  $M^\ast_1$  would result in the significant decrease of the timescale and luminosity in the second episodes. For the case of $M^\ast_1=0.2 M_\odot$, we find that the timescale of the second episode is shorter than 20 seconds\footnote{A low viscosity would increase the accretion time (e.g., Metzger et al. 2008) . In our calculations, we take $\alpha_1=\alpha_2=0.01$} and the luminosity is roughly 1 order of magnitude lower than that of the case $M^\ast_1=0.8 M_\odot$, as shown in Figure 3. Our results indicate that a bright EE component may be only detected for SGRBs with a massive disk.

\section{Conclusions and discussion}
We have proposed that both the outward angular momentum transfer and the switch of the magnetic barrier of a rotating BH-neutrino-cooled disk system may result in a long-lasting, impulse engine to produce several radiation episodes as observed in Type I GRBs. Based on the NDAF model, we have presented detailed calculations for some typical GRBs with detection of the EE component. Our numerical results in reasonable parameter sets well agree with the data.

Observationally, about half of the Type I GRBs have EE component detection with {\em Swift}/BAT. This may be due to both the physical and instrumental effects. Our result suggests that a highly magnetic, massive disk is required to produce an intense, long-lasting EE component. Massive disk may
from coalescence of a BH with a massive NS, i.e., $M_{\rm NS}\sim 2$ $M_\odot$ (Morrison et al. 2004; Dai et al. 2006; Demorest et al. 2010; Li et al. 2012). Klu\'{z}niak \& Lee (1998) showed that the mass of the disk from merger of a BH-NS binary may be larger than $0.5$ $M_\odot$ (see also Janka et al. 1999). For typical GRBs with EE detection by {\em Swift}/BAT, our model suggests that the disk should be  $\sim 0.8$ $M_\odot$. In case of the mass being lower than $\sim 0.2$ $M_\odot$, the EE may cannot be detected with {\em Swift}/BAT and its timescale is shorter than $20$ seconds. Both analytical and simulated investigations for coalescence of compact objects have been extensively studied (e.g., Ruffert et al. 1997; Klu\'zniak \& Lee 1998; Lee \& Klu\'zniak 1999; Ruffert 1999; Lee \& Ramirez-Ruiz 2002; Lee et al. 2005; Proga \& Zhang 2006; Ruffert \& Janka 2010). The suggested mass of the disk by these authors is much lower than that predicted by our model. Considering that the mass of a NS is $1-1.4$ $M_\odot$, Ruffert \& Janka (1997) pointed out that the coalescence of the NS-NS binary might result in a BH $\sim 2.5$ $ M_\odot$ surrounding by a disk with mass $\sim 0.1-0.2$ $M_\odot$. The key ingredients in our model are a massive disk and the angular momentum transfer, which may significantly prolong the lifetime of the GRB central engine for the EE component.  Therefore, the EE component may be a probe for merger process and formation of massive disk in the central engine of SGRBs.

The detection or not of the EE component might be also due to instrumental selection effect. Our model suggests that a disk with $M^\ast_1<0.2 M_\odot$ would not produce an EE component that can be detectable with {\em Swift}/BAT-like instruments. Assuming a constant radiation efficiency, the observed luminosity is proportional to the neutrino annihilation luminosity. Since the peak energy of the $\nu f_{\nu}$ spectrum is tightly correlated with the radiation luminosity (Liang et al. 2004; Yonetoku et al. 2004), the EE component of some SGRBs may be out of the instrument bands and only the initial hard-short episode can be detectable as usually seen in typical SGRBs. With the results for $M^\ast_1=0.8 M_\odot$ and $M^\ast_1=0.2 M_\odot$ one can observed a luminosity-duration correlation for the EE component. This seems to be at odds with observations that show the opposite trend: longer events have a lower luminosity. Note that the observations for the EE component greatly suffer from the selection effect of instrumental sensitivity. We check the observed luminosity-duration correlation for both the initial hard spikes and the EE components for the GRBs with EE detection reported in Zhang et al. (2009), but no statistically accepted correlation is found. The relation between the luminosity and duration of our model is parameter-dependent. This calls for a detailed analysis based on Monte Carlo simulations in order to elaborate this relation and its observational biases.

Some caveats for our model should be discussed. The neutrino annihilation luminosity is correlated positively with the accretion rate. In order to ensure the mass of NS less than about $2$ $M_\odot$, a moderate accretion rate is required in our model. The derived accretion rates for both the initial and late episodes from our model are less than $0.05$ $M_\odot$ s$^{-1}$, which is lower than the typical ones, i.e., $0.1\sim 1 M_\odot$ s$^{-1}$. Note that the typical values are for a spatially uniform neutrino annihilation luminosity. In our calculations, the spatial distribution of the neutrino annihilation luminosity is collimated (e.g., Popham et al. 1999; Liu et al. 2007). The jet opening angle is conservatively assumed to be $10^\circ$ in our calculations\footnote{The jet opening angle of SGRBs may be even much smaller than $10^\circ$, such as $\sim 0.3^\circ$ for GRB 090510 (Corsi et al. 2010; He et al. 2011; Fan \& Wei 2011)}. Therefore, the isotropic-equivalent accretion rate would be larger than $1 \sim 2$ order of magnitude than our results, which is consistent with the typical values used by previous authors.

We have considered the switch of magnetic field for the accretion processes, but the effects to confine the jet opening angle by the BZ mechanism (Blandford \& Znajek 1977) and the enhance of the neutrino annihilation are ignored. These effects may also greatly impact on the jet luminosity. Lei et al. (2009) investigated the magnetic coupling between the BH and disk. They found that the luminosity of neutrino annihilation is much larger than the luminosity of NDAF without magnetic field. Recently, Barkov \& Pozanenko (2011) proposed a two jet model which describes both main and EE components by different off-axis position of observer. Their model involves a short duration jet powered by heating due to neutrino annihilation and a long-lived BZ jet with significantly narrow opening angle. The BZ mechanism can replace the neutrino annihilation to produce the main emission and EE in our model. Since the lack of information of the intensity and distribution of the magnetic field, we restrict to discuss the neutrino annihilation as the main power source in our model.

\section*{Acknowledgments}
We thank W{\l}odzimierz Klu\'{z}niak, Bing Zhang, Feng Yuan, Ye-Fei Yuan, Yi-Zhong Fan, Li Xue, Xiao-Hong Zhao, and Da-Bin Lin for beneficial discussions and comments. This work is partially supported by the National Basic Research Program (``973'' Program) of China (Grant 2009CB824800), the National Natural Science Foundation of China (Grants 10873005, 10873009, 11003004, 11025313, 11033002, 11073015, 11103015, 11222328 and 11233006), and Guangxi Science Foundation (2010GXNSFC013011 and special support with Contract No. 2011-135).

\clearpage

\begin{figure}
\centering
\includegraphics[width=0.8\textwidth]{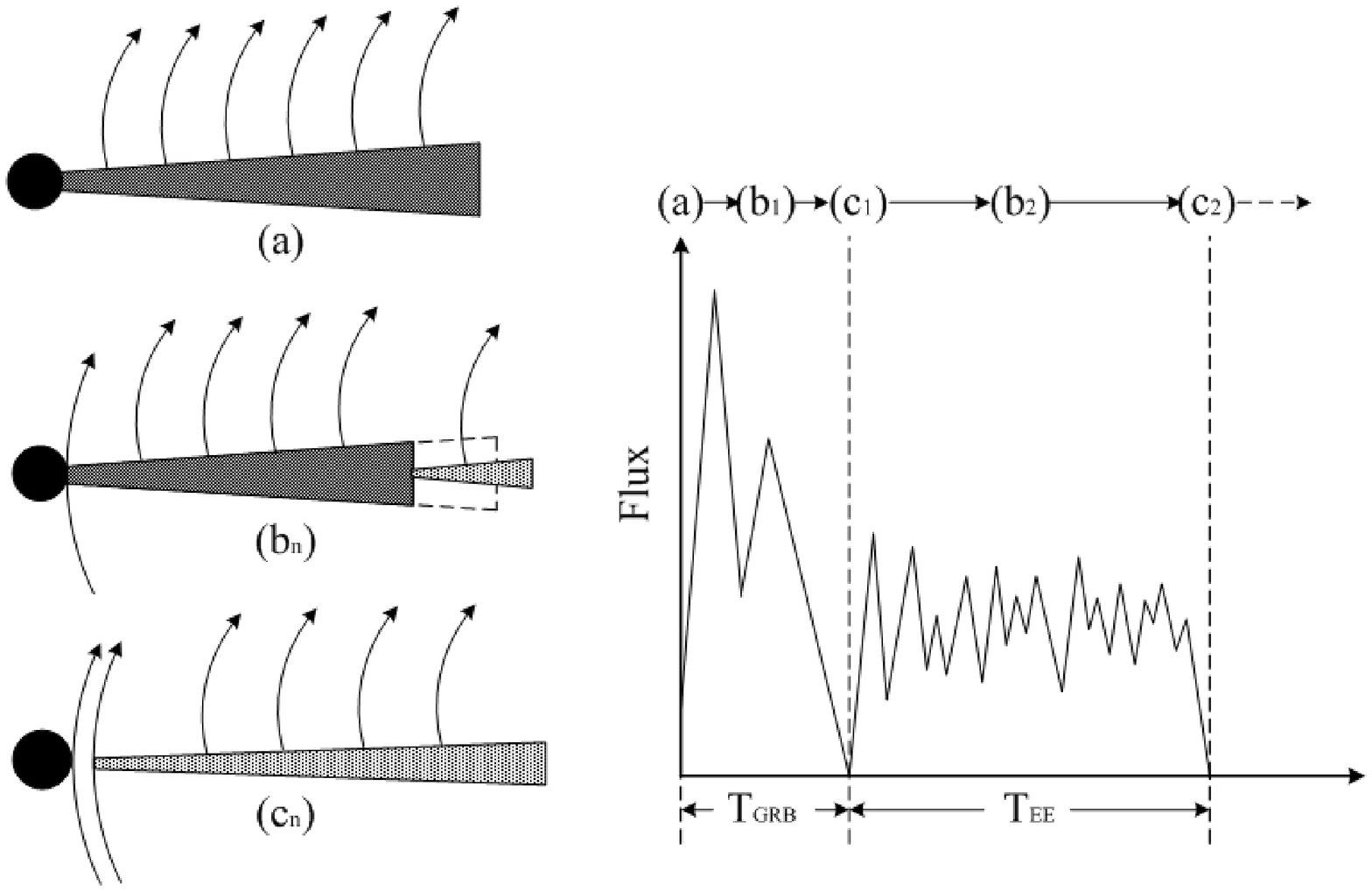}
\caption{Schematic illustration of our model and corresponding cartoon lightcurve: (a) the initial state of the central engine---the filled circle stands for the BH and the fuscous trapeziform region for the disk with magnetic field (curves); $(b_{\rm n})$ angular momentum transfer process and outward flow (light gray region); $(c_{\rm n})$ magnetic barrier in vicinity of the BH, where $n$ is for the $nth$ emission episode.} \label{figure1}
\centering
\end{figure}

\clearpage

\begin{figure}
\centering
\includegraphics[width=0.4\textwidth]{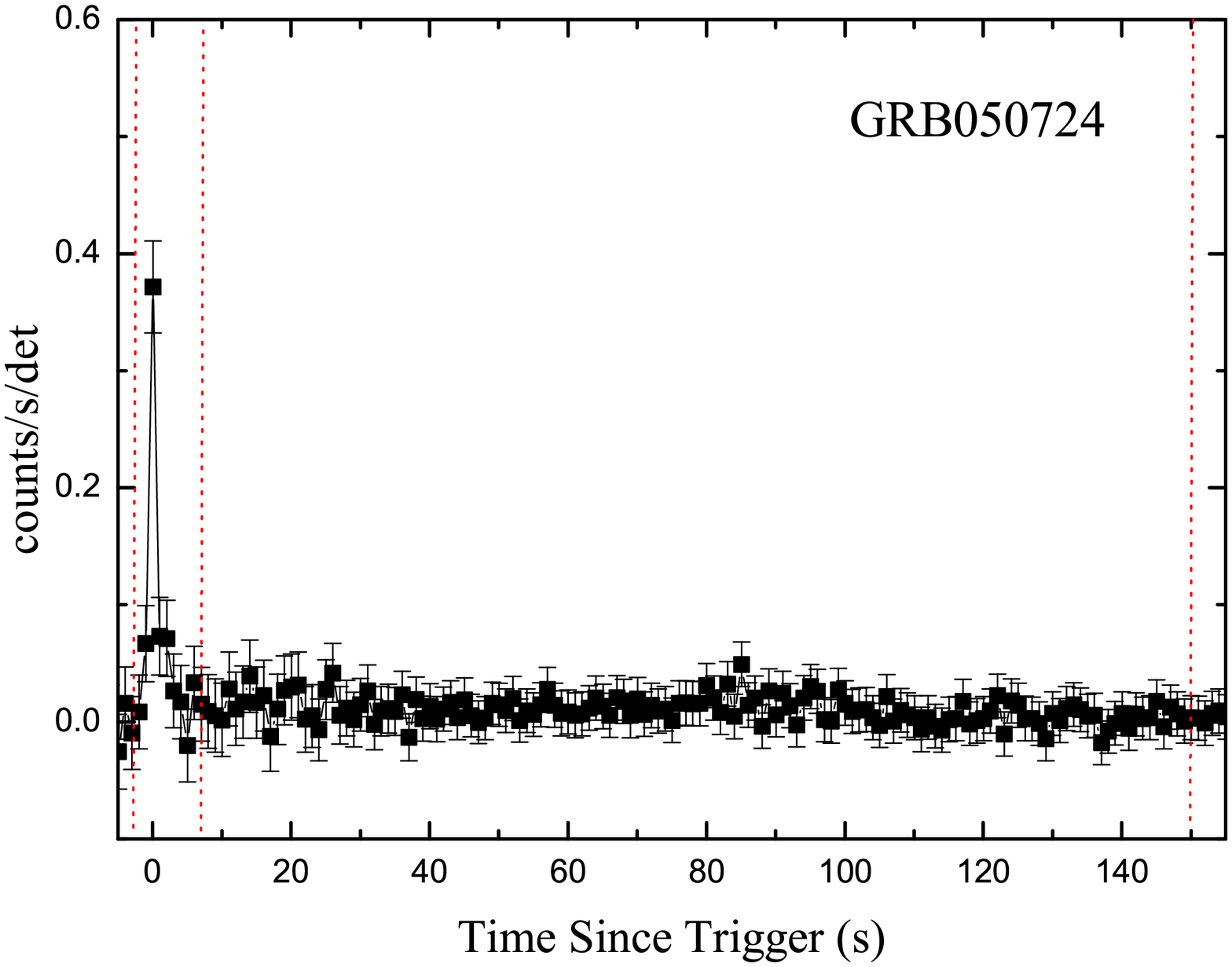}
\includegraphics[width=0.4\textwidth]{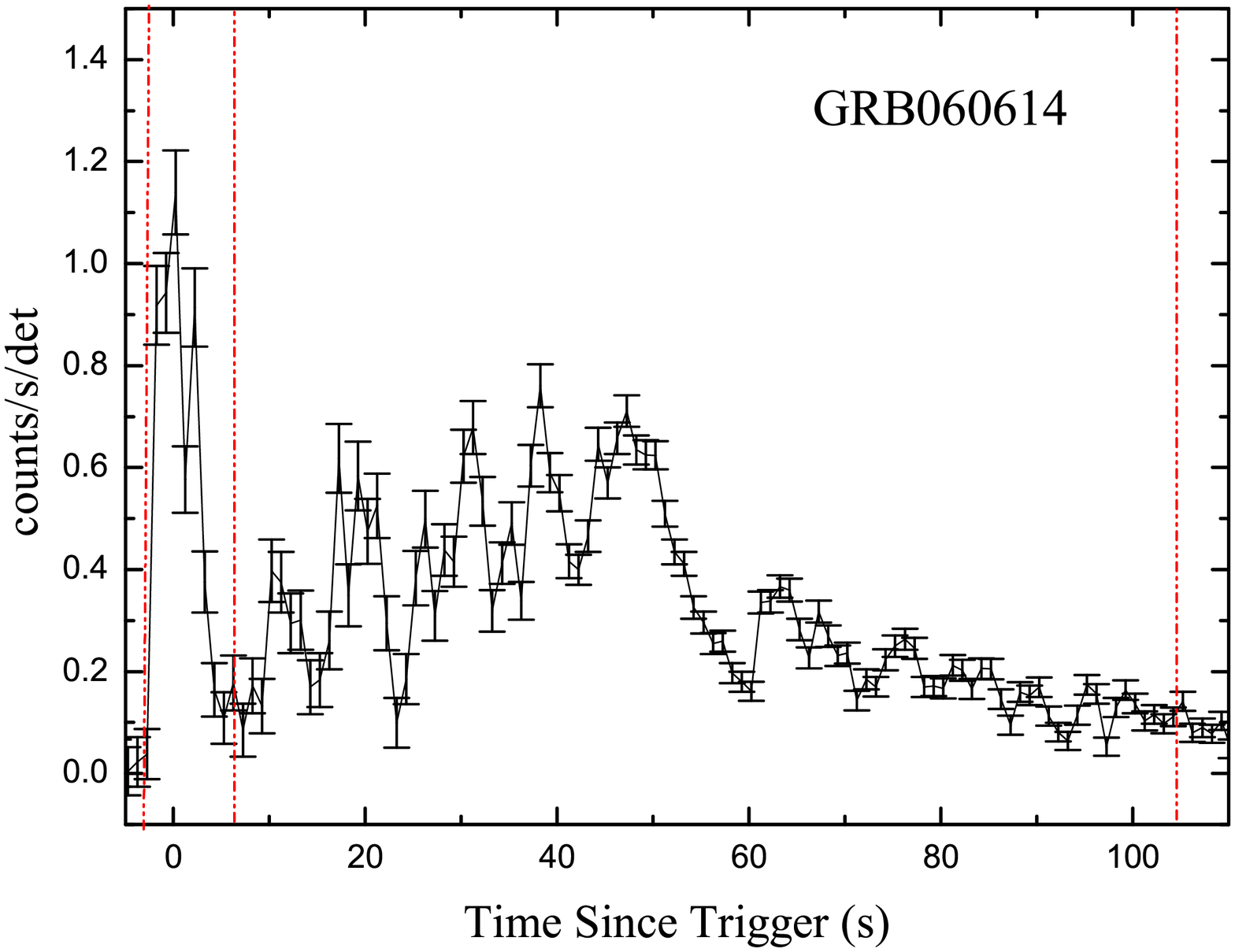}
\includegraphics[width=0.4\textwidth]{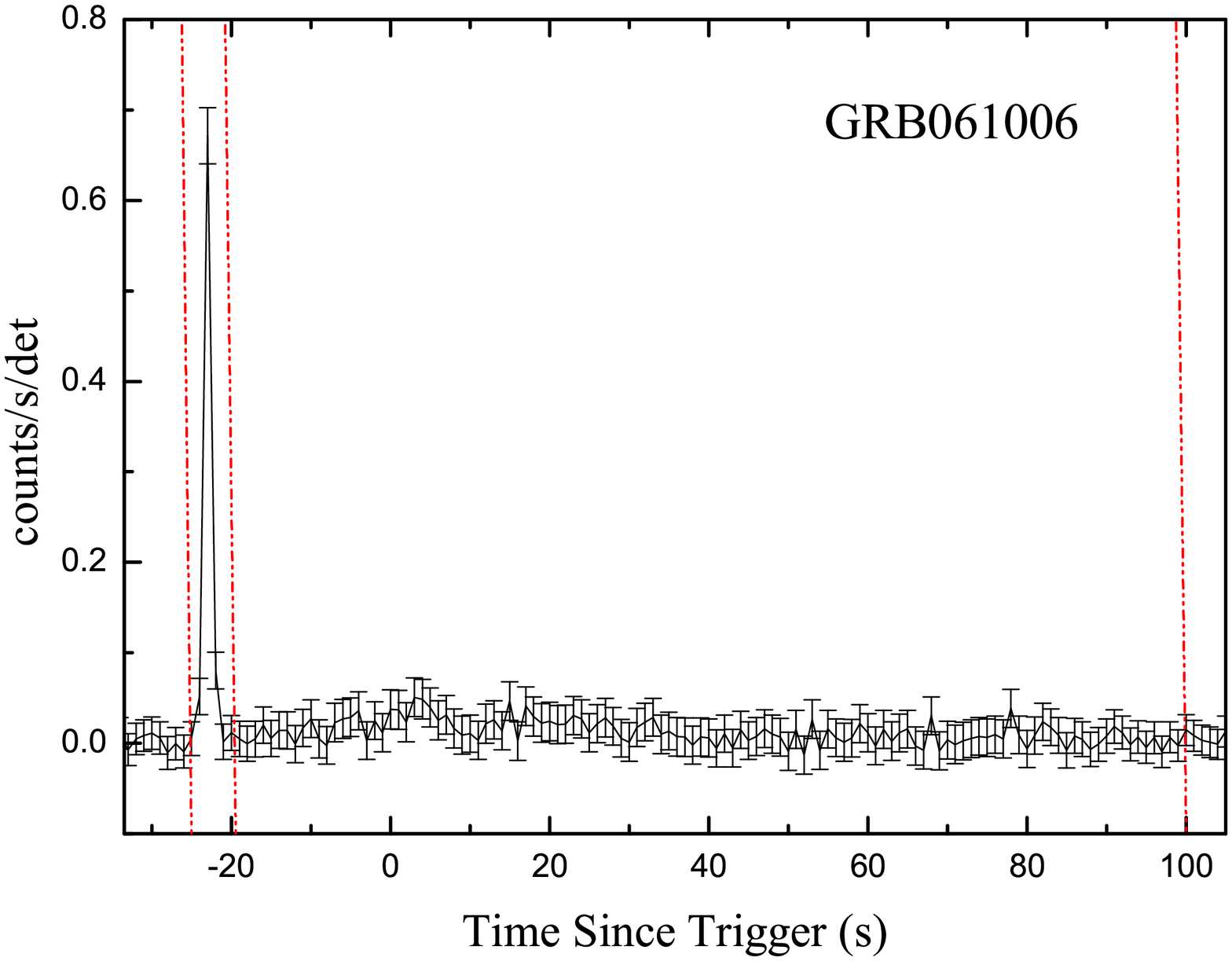}
\includegraphics[width=0.4\textwidth]{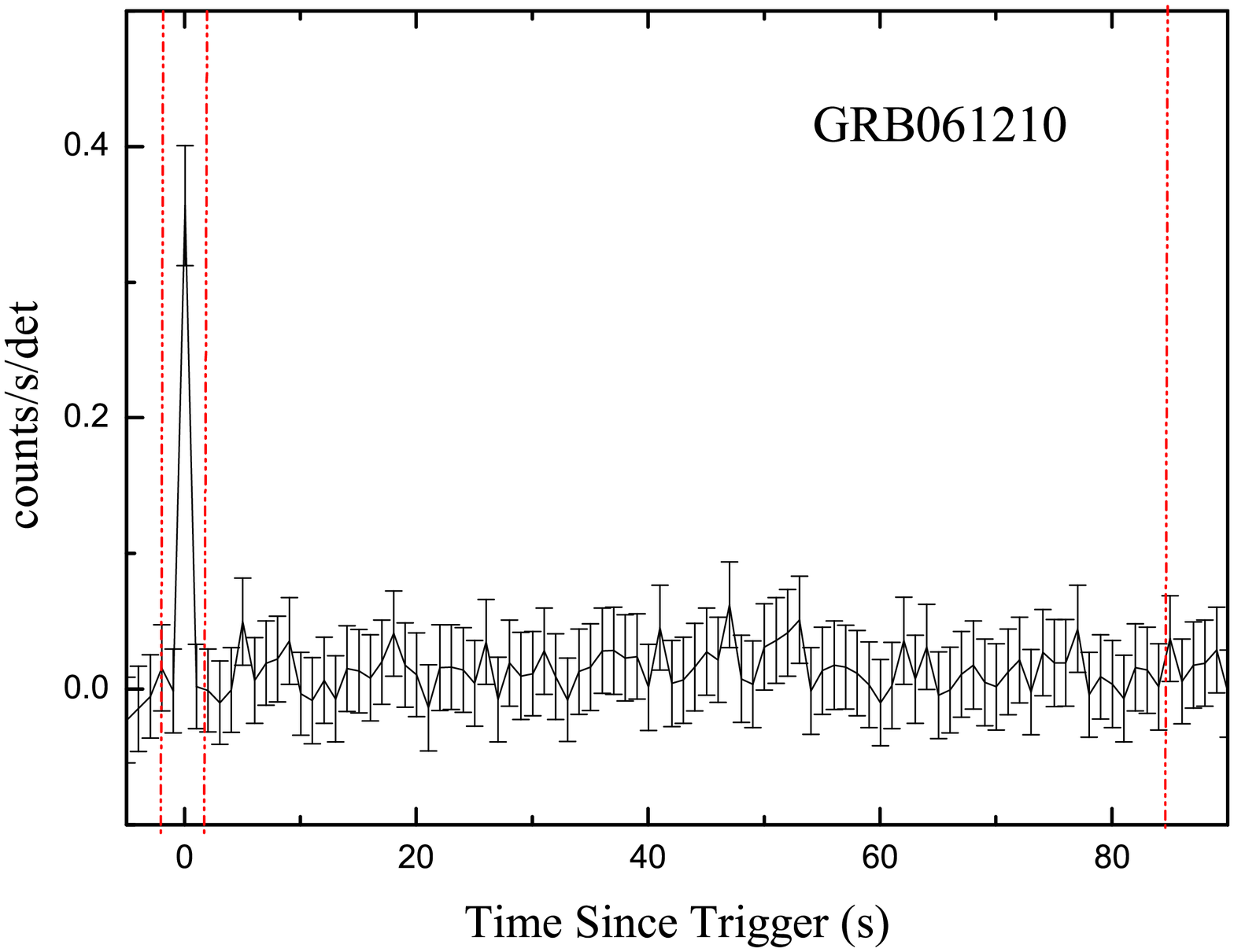}
\includegraphics[width=0.4\textwidth]{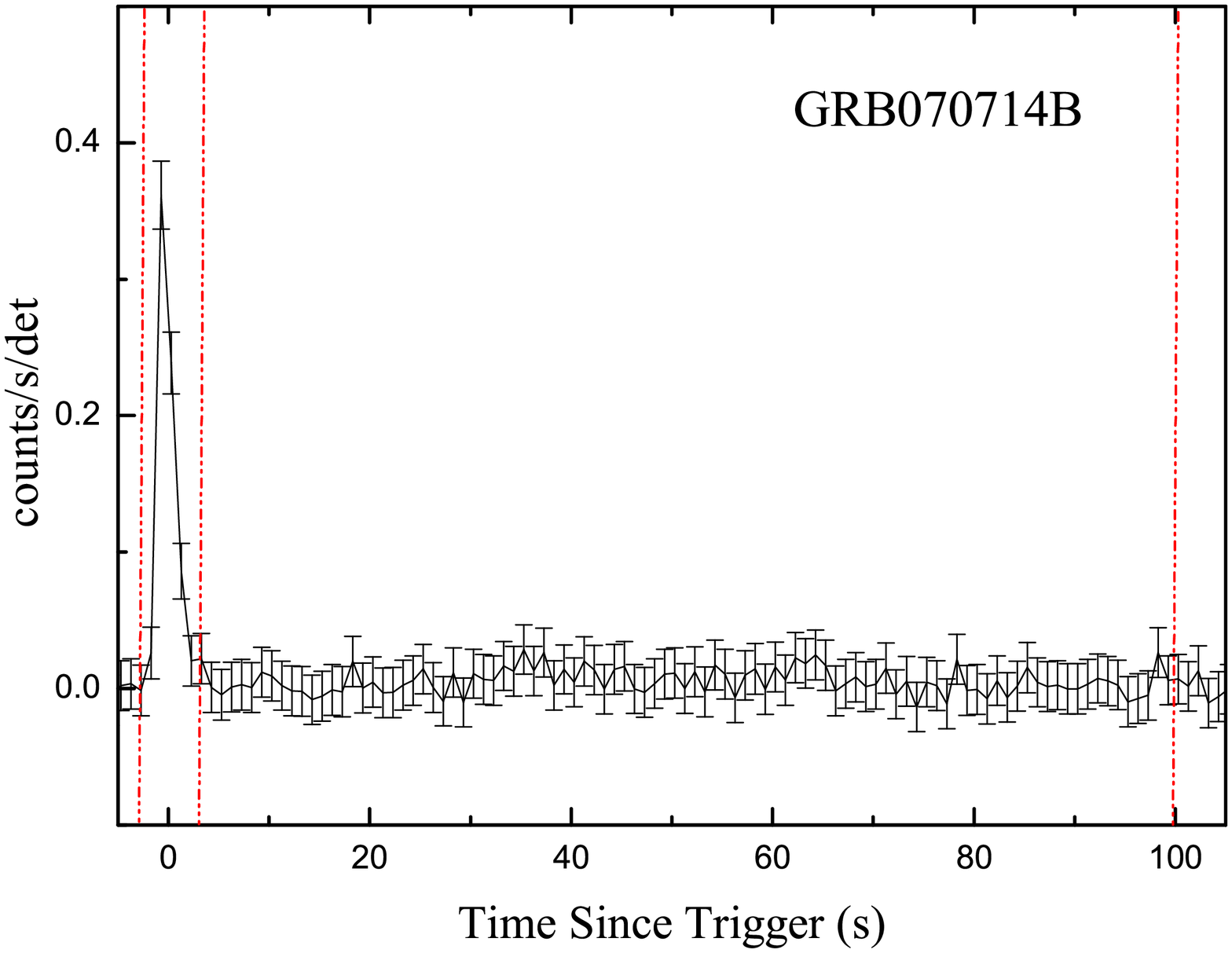}
\includegraphics[width=0.4\textwidth]{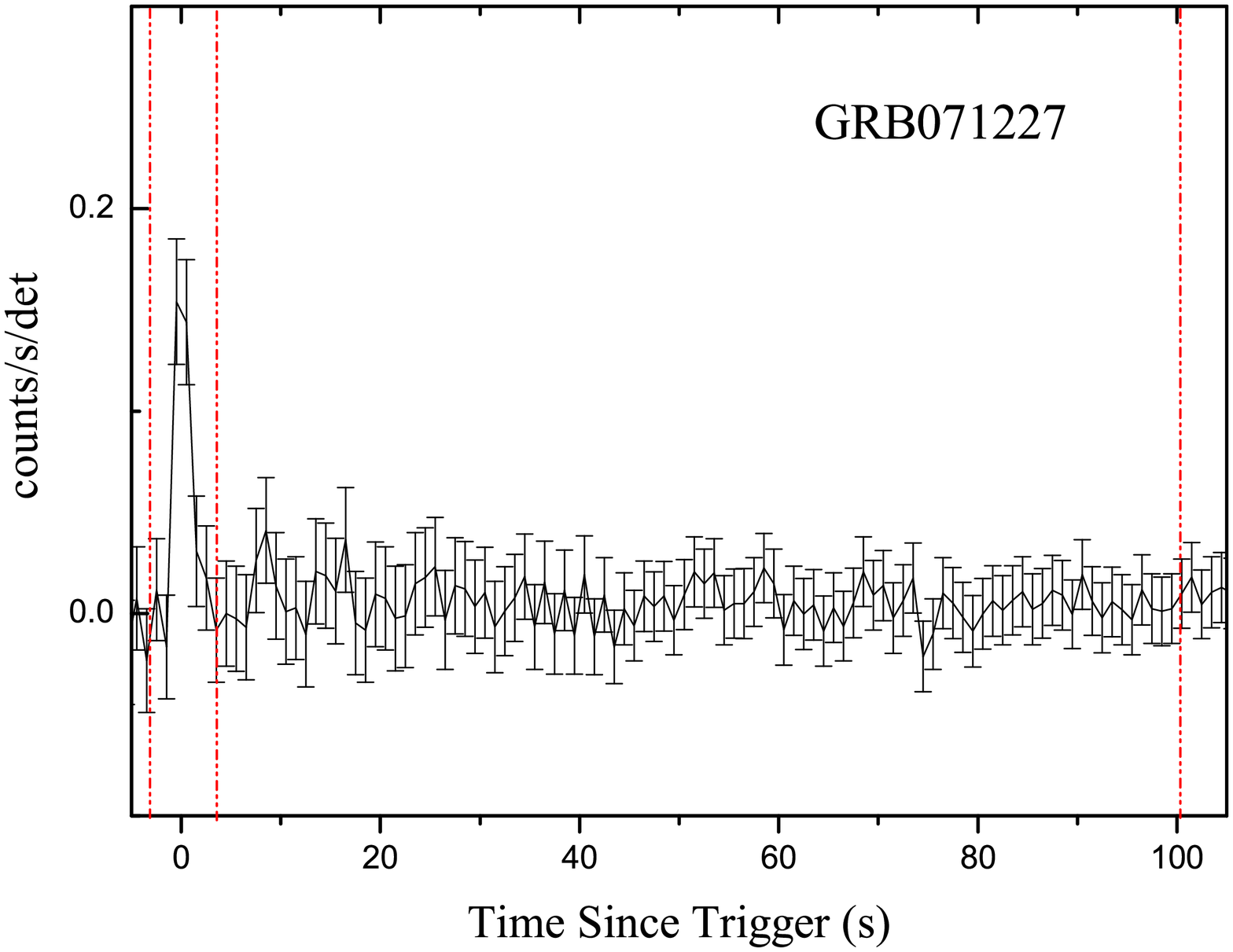}
\caption{BAT lightcurves of six SGRBs with detection of the EE component.}
\label{figure2}
\centering
\end{figure}

\begin{figure}
\centering
\includegraphics[width=0.8\textwidth]{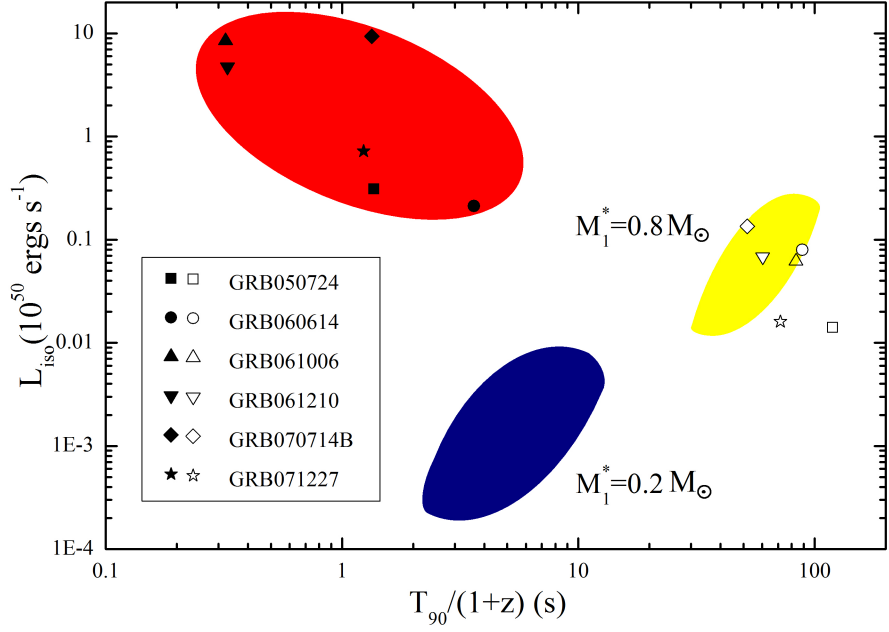}
\caption{Our numerical results for the six SGRBs shown in Figure 2. The filled symbols and open symbols figure the main and the corresponding EE components of the SGRBs. The first episodes  of these GRBs are in the left-top circle of the figure. We correspondingly derive a region of the second episodes with the universal parameters for $M^\ast_1=0.8 M_\odot$. Besides GRB 050724 and 071227, the second episodes of the other four GRBs are in the region. The EE component of the two GRBs are likely a late flare, which may have different physical origin as mentioned in \S 1. The corresponding region for $M^\ast_1=0.2 M_\odot$ is also shown. It indicates that the luminosity of the second episode is roughly 1 order of magnitude lower than that for the case of $M^\ast_1=0.8 M_\odot$ and the timescale is shorter than 20 seconds.The switch timescale of the magnetic barrier is 0.1-1 seconds. This timescale is much smaller than the observed timescale of the EE component. We thus ignore it here.}
\label{figure3}
\centering
\end{figure}

\clearpage

\end{document}